# Ultra-high-resolution imaging of moiré lattices and superstructures using scanning microwave impedance microscopy under ambient conditions

Kyunghoon Lee[1,2,3,†], M. Iqbal Bakti Utama[1,2,4,†], Salman Kahn[1,2,†], Appalakondaiah Samudrala[5], Nicolas Leconte[5], Birui Yang[1], Shuopei Wang[6], Kenji Watanabe[7], Takashi Taniguchi[8], Guangyu Zhang[6], Alexander Weber-Bargioni[9], Michael Crommie[1,2,3], Paul D. Ashby[9], Jeil Jung[5], Feng Wang[1,2,3,*], Alex Zettl[1,2,3,*]

[1] Department of Physics, University of California at Berkeley, Berkeley, California 94720, USA
[2] Materials Sciences Division, Lawrence Berkeley National Laboratory, Berkeley, California, 94720, USA
[3] Kavli Energy NanoSciences Institute at the University of California, Berkeley and the Lawrence Berkeley National Laboratory, Berkeley, California 94720, United States
[4] Department of Materials Science and Engineering, University of California at Berkeley, Berkeley, California 94720, USA
[5] Department of Physics, University of Seoul, Seoul, South Korea
[6] Beijing National Laboratory for Condensed Matter Physics and Institute of Physics, Chinese Academy of Sciences, Beijing 100190, China and Collaborative Innovation Center of Quantum Matter, Beijing, China
[7] Research Center for Functional Materials, National Institute for Materials Science, 1-1 Namiki, Tsukuba, 305-0044, Japan
[8] International Center for Materials Nanoarchitectonics, National Institute for Materials Science, 1-1 Namiki, Tsukuba, 305-0044, Japan
[9] Molecular Foundry, Lawrence Berkeley National Laboratory, Berkeley, California, 94720, USA
[†] These authors contributed equally to this work
[*] E-mail: azettl@berkeley.edu; fengwang76@berkeley.edu

**Abstract**

Two-dimensional heterostructures with layers of slightly different lattice vectors exhibit a new periodic structure known as moiré lattices. Moiré lattice formation provides a powerful new way to engineer the electronic structure of two-dimensional materials for realizing novel correlated and topological phenomena. In addition, superstructures of moiré lattices can emerge from multiple misaligned lattice vectors or inhomogeneous strain distribution, which offers an extra degree of freedom in the electronic band structure design. High-resolution imaging of the moiré lattices and superstructures is critical for quantitative understanding of emerging moiré physics. Here we report the nanoscale imaging of moiré lattices and superstructures in various graphene-based samples under ambient conditions using an ultra-high-resolution implementation of scanning microwave impedance microscopy. We show that, quite remarkably, although the scanning probe tip has a gross radius of ~100 nm, an ultra-high spatial resolution in local conductivity profiles better than 5 nm can be achieved. This resolution enhancement not only enables to directly visualize the moiré lattices in magic-angle twisted double bilayer graphene and composite super-moiré lattices, but also allows design path toward artificial synthesis of novel moiré superstructures such as the Kagome moiré from the interplay and the supermodulation between twisted graphene and hexagonal boron nitride layers.

Moiré lattices with large periodicity can be realized in two-dimensional (2D) heterostructures composed of atomically thin layers with slightly different lattice vectors, either due to a lattice mismatch or a small-angle twist. Such moiré lattices generate a new length and energy scale in stacked 2D materials and provide an exciting new platform to engineer novel correlated phenomena and topological physics in van der Waals heterostructures (*1-12*). Superstructures of moiré lattices can emerge if multiple layers with similar lattice vectors are stacked together, offering extra flexibility to design novel quantum phenomena.

Quantitative characterization of the moiré lattice and superstructure in a device configuration is critical for understanding and controlling the rich moiré physics in 2D heterostructures. Traditionally, the structure of moiré lattices is imaged with transmission electron microscopy (TEM) (*13, 14*) and scanning tunneling microscopy (STM) (*15-20*), but these methods have low throughput and require specialized sample preparation that is largely unsuitable for functional devices. Different imaging modalities based on atomic force microscopy (AFM) have also been explored to study moiré lattices (*21-26*), but imaging moiré lattices and superstructures with sufficient sensitivity and resolution is challenging.

Relative to other AFM- and TEM-based techniques, scanning microwave impedance microscopy (sMIM) (*27*) is an attractive moiré imaging tool as since, in principle, it combines the benefit of reasonable spatial resolution, high sensitivity to local electrical properties, and compatibility with functional electronic devices. Indeed, we here demonstrate an ultra-high-resolution implementation of sMIM, which we term uMIM, with which we perform nanoscale imaging of moiré lattices and superstructures in various graphene-based devices under ambient conditions. Remarkably, we show that a spatial resolution better than 5 nm can be achieved with uMIM despite the gross scanning probe tip radius being ~100 nm. Owing to the strong sensitivity of the uMIM signal on the local conductivity of the sample, uMIM can probe not only the regular moiré lattice in twisted graphene and graphene/hBN heterostructures, but also the moiré superstructures formed by multiple stacked layers. We use uMIM imaging to reveal several moiré superstructures, including a supermodulation of the moiré lattice and a novel Kagome-like moiré structure that arises from the interplay between closely aligned twisted graphene and hBN layers. Such moiré superstructures can offer new avenues to engineer quantum phenomena in van der Waals heterostructures.

A highly simplified schematic for our measurements is illustrated in Fig. 1A. Scanning microwave impedance microscopy probes the local complex tip-sample admittance and the signals are represented by the real and imaginary responses (*27*). The tip-sample admittance depends sensitively on the local sample conductivity. Fig. 1B shows the calculated real and imaginary uMIM signals, uMIM-Re and uMIM-Im, as functions of the sample sheet conductance ($g$), using the inferred ultra-high-resolution modified tip geometry (see Methods and Supplementary Materials). We note that the uMIM-Im signal is informative for a rapid assessment of the local conductivity because it increases monotonically with $g$. sMIM-Re and sMIM-Im for the gross (unmodified) tip geometry (scanning electron microscope (SEM) image of Fig. 1A), are presented in fig. S10A.

Scanning microwave impedance microscopy can be considered a microwave version of the apertureless near-field optical microscopy. However, unlike typical near field microscopy, our experiments are performed in contact mode where the tip-sample distance is maintained in the repulsive regime. Consequently, the electromagnetic coupling between the tip and sample is highly localized at the tip apex. Previous studies have demonstrated that the resolution of sMIM and other near-field techniques is limited by tip geometry, with the tip radius setting the resolution limit (typically 50-100 nm). To achieve ultra-high-resolution, we use a conventional sMIM instrument in air at room temperature, but specially prepare (*i.e.*, condition) the tip and sample. Briefly, using the same tip throughout, the sample is first pre-scanned over a large area. A detailed small-area data-collecting scan is then performed. We hypothesize that the conditioning step results in: (1) sample surface cleaning, and (2) tip geometry modification. Regarding the latter, both electrically conducting and insulating adsorbates from the sample redistribute onto the tip apex eventually forming a protruding thin conducting chain supported by an insulating matrix. This result in an electromagnetic coupling that is dominated by the metallic chain, while the signal contribution from the remaining part of the bulk recessed tip is suppressed. This scenario, somewhat akin to the enhanced resolution afforded to STM-AFM by tip-adsorbed CO molecules (*28, 29*) and to conductive AFM by the formation of metallic filament near tip apex (*30*), facilitates spatial resolution significantly finer than the gross radius of the tip (see Methods, figs. 2 and 11 in the Supplementary Materials for further illustrations and discussion) (*31*).

We first demonstrate the capability of uMIM by imaging the moiré superlattice in twisted double bilayer graphene (tDBG). Similar to twisted bilayer graphene (tBG) (*2*), tDBG contains a continuously varying stacking sequence within the moiré unit cell (*6*). The tDBG unit cell contains one truncated triangular domain each of ABAB (Bernal) and ABCA (rhombohedral) stacking, with a nearly circular ABBC stacking at the vertices of each triangle (Fig. 1C). Fig. 1D shows a uMIM map of a tDBG sample where we resolve the moiré period from the contrasts of signal arising from the different local stacking. The measured moiré period of this sample is $(10.9 \pm 0.7)$ nm, corresponding to a twist angle of $\theta = (1.3 \pm 0.1)°$, the "magic angle" in tDBG for which field-tunable correlated insulating and superconducting states have been reported (*7-10*).

We clearly resolve three different domains in the tDBG moiré lattice with distinct uMIM signals (Fig. 1D-E). Our result demonstrates the usefulness of uMIM in identifying fine structures of moiré lattices in 2D heterostructures based on the local conductivity. We assign the domains with the weakest uMIM-Im signal (*i.e.*, the smallest local conductivity) as the ABBC stacking. Analogous to the circular AA stacking in tBG (*17*), the circular ABBC domains in tDBG occupies a proportion of the moiré unit cell that is relatively larger than the other two stacking domains. However, due to the lattice relaxation with decreasing twist angle at $\theta < 1°$, the proportion of ABBC domain in the moiré unit cell continually decreases up until it disappears for near-0° twist, leaving only triangular ABAB-ABCA domains (*e.g.*, Fig. 2C). Consistent with this trend, we observe the decrease of the size proportion in domains with the weakest uMIM-Im signal in tDBG with twist angle of 0.6° (fig. S4). Meanwhile, the two different contrasts in regions between ABBC stacking are attributed to the ABCA and ABAB domains.

The spatial resolution capability of uMIM is further demonstrated in the imaging of "moiré defects", which interrupt the long-range periodicity of the moiré lattice. In Figs. 1F-G (and fig. S5), we can readily resolve the moiré defects with sub-5 nm resolution, outperforming other optical near-field microscopes.

Fig. 2 demonstrates the universal applicability of uMIM in resolving moiré structures in various graphene-based systems. Particularly, uMIM can visualize the inhomogeneity and defect structures in the moiré lattices (also see figs. S7-8). Fig. 2A shows the moiré observed in epitaxially-grown monolayer graphene/hBN. Such aligned graphene/hBN system exhibits a transition into a commensurate moiré stacking (*21*), where the graphene lattice expands to accommodate the mismatch with hBN. This results in a hexagonal domain-wall network (thin bright lines in Fig. 2A) with accumulated strain. The moiré period from Fast-Fourier transform (FFT, Fig. 2A inset) is determined to be (16.1 ± 0.5) nm, close to that expected for perfectly aligned graphene and hBN layers.

At near-0° twist, the lattice energetically prefers to reconstruct into relaxed triangular domains with an alternating stacking sequence (*14, 32*). In twisted trilayer graphene (tTG) and tetralayer tDBG, these domains correspond to Bernal (ABA and ABAB) and rhombohedral stacking (ABC and ABCA), respectively. Each stacking order results in different low-energy electronic band structure (*33*). Fascinatingly, the rhombohedral stacking in few-layer graphene owns flat bands with bandwidth that scales inversely with the layer number (*33*), making it an attractive platform to study correlation physics (*5, 6, 11, 12*). With the rarity of rhombohedral stacks from their relative instability, the near-0° twist presents an alternative route for generating the stacking (*6*).

The low Fermi velocity in rhombohedral stacking as compared to Bernal stacking should result in a lower electrical conductivity for the rhombohedral domain. Indeed, in Fig. 2B and 2C, the triangular domains in tTG and tDBG are successfully resolved owing to this local conductivity variation, where the domains with lower uMIM-Im signal are assigned to the rhombohedral stack (ABC in tTG, ABCA in tDBG). In addition, Fig. 2B also shows possible strain effects at the domain walls and vortices, leading to a deviation of the domain shape from regular triangle.

In addition to the conventional moiré lattices, our ultra-high sensitivity uMIM allows for imaging of moiré superstructures from three underlying lattices with different lattice vectors.

Although heterostructures with multiple coexisting moiré has been imaged before with AFM (*25*), the resulting moiré superstucture has not been observed in ambient conditions previously. Fig. 3A-C shows a BG/BG/hBN heterostructure where the two bilayer graphenes (BGs) are slightly twisted and the bottom BG is nearly aligned with the underlying hBN flake. Two moiré lattices are present: one from the tDBG and the other from the aligned BG/hBN. They can interfere and form a supermodulation pattern with very large periodicity. Fig. 3A shows such moiré superstructure. The FFT pattern reveals three distinct periods (Fig. 3B). Two sets of the moiré periods (red and blue dashed hexagons) have similar periodicity of (14.1 ± 0.4) nm and (12.3 ± 0.7) nm, and we identify them as the underlying moiré lattices between BG/BG and BG/hBN (*i.e.*, $\lambda_{BG/BG}$ and $\lambda_{BG/hBN}$, respectively). The other set of spots in FFT (purple dashed hexagon) has a much longer period of 45.7 nm, which we ascribe to the supermodulation of the moirés. Such a periodicity cannot be assigned to BG/hBN moiré as it requires too large a strain in the graphene lattice, nor to BG/BG moiré, as at this length scale the moiré should relax into triangular domains (*14*). As a check, we compare this periodicity with the expected period of a super-moiré from the two constituent moirés (*34*):

$$\lambda_s = \frac{1}{\sqrt{\left(\frac{1}{\lambda_{BG/BG}}\right)^2 + \left(\frac{1}{\lambda_{BG/hBN}}\right)^2 - \frac{2\cos(\Delta\theta_s)}{\lambda_{BG/BG}\,\lambda_{BG/hBN}}}} \quad (1)$$

where $\Delta\theta_s = 14.7°$ is the misorientation between the moirés. This results in $\lambda_s$ of ~45.2 nm, which agrees well with the observed periodicity of 45.7 nm. Fig. 3C shows the real-space image for each moiré as filtered from the first-order spots in the FFT.

A different moiré superstructure can exist when a near-0° tDBG with triangular relaxed domains are intertwined with the commensurate moiré from aligned bottom BG/hBN (Fig. 3D). This combination results in an apparent reconstruction of the BG/hBN moiré near the ABAB-ABCA domain walls. An indication of this is seen in the FFT (Fig. 3E), where the pattern corresponding to the BG/hBN moiré (dashed blue square) no longer appear as single isolated spots.

Filtering the uMIM-Im image by selecting only these features (Fig. 3F) clarifies the supermodulation of BG/hBN moiré occurring near the domain wall. We believe that this feature is not an artifact, as similar behavior can be seen in a smaller area uMIM-Im scan (Fig. 3F) and from the topographic image (fig. S9). Such modification of the moiré structure might lead to a modified electronic spectrum, which may need to be incorporated in theoretical calculations of the electronic structure.

uMIM imaging allows us to readily explore other moiré superstructures with desirable physical properties. A special example is the Kagome lattice, which has attracted significant attention as a platform to study Hubbard physics due to the presence of flat bands and exotic quantum and magnetic phases with non-trivial ordering (*35-37*). However, crystals with a natural Kagome lattice are relatively rare. In ultracold atoms research, a Kagome lattice can be simulated via an optical superlattice. For example, one type of such a corner-sharing triangular superlattice, called the trimerized Kagome lattice, can be constructed by superimposing two triangular optical lattices with 2:1 periodicity (*38, 39*).

Inspired by the artificial Kagome lattice in optical superlattices, we construct a solid state Kagome-like moiré superlattice in BG/BG/hBN systems by superimposing $\lambda_{BG/BG}:\lambda_{BG/hBN} = 2:1$ and maintaining azimuthal alignment of the moiré constituents. These are achieved by setting the twist angle of both the BG/BG and BG/hBN interfaces to $\theta \approx 0.6°$ (Fig. 4A-B). We visualize such a special moiré composite with uMIM (Fig. 4C) and confirm from FFT the presence of two aligned moirés with periods of $\lambda_{BG/BG} = (24.5 \pm 1.1)$ nm and $\lambda_{BG/hBN} = (13.1 \pm 0.1)$ nm. The resulting structure is examined in more detail in Fig. 4F, and compared with the expected structure for an ideal Kagome lattice (Fig. 4G).

The construction of a solid state Kagome-like lattice based on moiré patterns is an interesting pathway to explore the possibility of generating nearly flat bands and host correlated phases in the vicinity of the Fermi level. While the superlattice miniband structures are susceptible to changes in the details of the atomic and electronic structure models, the band flatness should further be enhanced since the bilayer graphene within the tDBG has less dispersive bands than a Dirac cone at the K points, and since a gap can open through interaction with the substrate. Fig. 4H shows the calculated band structure around the Fermi level of the Kagome-like moiré (see Supplementary Materials for calculation details). A series of minibands with small bandwidth are observed, including flat conduction bands with a bandwidth of ~3 meV.

In summary, we have demonstrated ultra-high-resolution uMIM as a facile, high-throughput, and noninvasive method for characterizing moiré lattices and superstructures, as well as moiré defects and the creation of tailored Kagome superlattices in multilayer stacks of van der Waals heterostructures. The enhanced capability enabled by uMIM should facilitate better understanding and heterostructure design paths for novel correlated quantum phenomena in more advanced moiré superstructures.

**Acknowledgements:**

The authors thank R.C. Chintala (PrimeNano Inc.) for technical assistance.

**Funding:**

This work was supported by the Director, Office of Science, Office of Basic Energy Sciences, Materials Sciences and Engineering Division, and Molecular Foundry of the US Department of Energy under contract no. DE-AC02-05-CH11231, primarily within the sp2-Bonded Materials Program (KC2207) which provided for development of the project concept and uMIM measurements. Additional support was provided by the National Science Foundation, under grant DMR-1807322, which provided for preliminary AFM topography measurements. The graphene device fabrication is supported as part of the Center for Novel Pathways to Quantum Coherence in Materials, an Energy Frontier Research Center funded by the U.S. Department of Energy, Office of Science, Basic Energy Sciences. K.W. and T.T. acknowledge support from the Elemental Strategy Initiative, conducted by the MEXT, Japan, Grant Number JPMXP0112101001, JSPS KAKENHI Grant Numbers JP20H00354 and the CREST (JPMJCR15F3), JST. A.S. was supported by the Korean National Research Foundation grant NRF-2020R1A2C3009142, and KISTI computational resources through grant KSC-2020-CRE-0072. N.L. acknowledges the Korean national Research Foundation grant NRF-2018R1C1B6004437, the Korea Research Fellowship Program funded by the Ministry of Science and ICT (KRF-2016H1D3A1023826), as well as the computational resources by KISTI through grant KSC-2018-CHA-0077. J.J. was supported by the Samsung Science and Technology Foundation under project SSTF-BAA1802-06.

**Author contributions:**

K.L. and M.I.B.U. conceived the idea of the project, performed measurements, analyzed the data with P.D.A and A.W-B., and wrote the manuscript with input from all authors. S.K., B.Y., K.L., and M.I.B.U. fabricated all samples and devices. A.S., N.L., and J.J. calculated the band structures. S.W. and G.Z. performed CVD growth of graphene. K.W. and T.T. grew the hBN single crystal. A.Z. and F.W. supervised the project.

**Competing interests:**

The authors declare that they have no competing interests.

**Data and materials availability:**

All data needed to evaluate the conclusions in the paper are present in the paper and/or the Supplementary Materials. Additional data related to this paper may be requested from the authors.

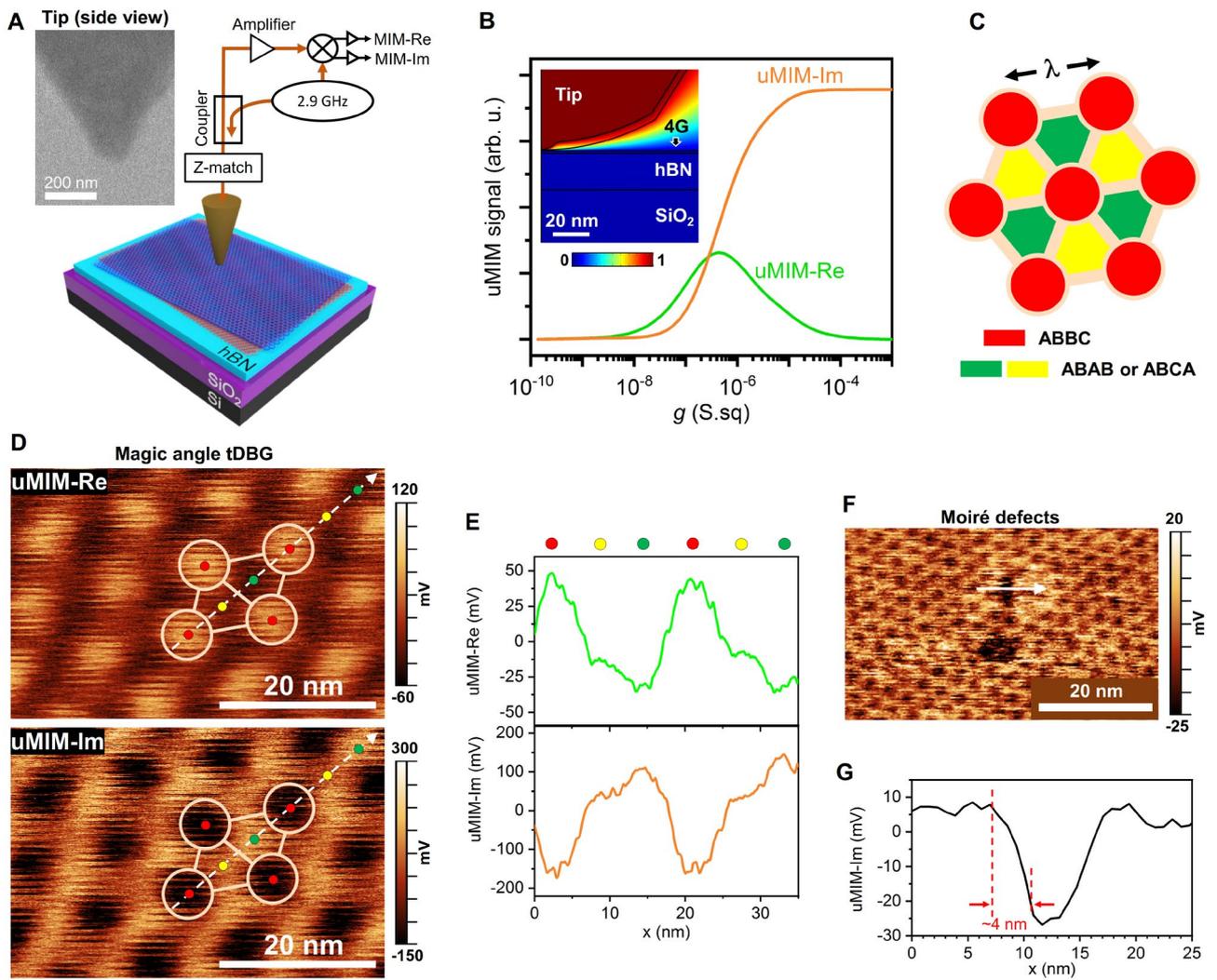

**Fig. 1. Imaging mechanism and spatial resolution of uMIM. (A)** Measurement configuration. *Inset*: Gross SEM image of the tip. **(B)** Calculated uMIM signals as a function of the sample sheet resistance, assuming modified tip (see Supplementary Materials). *Inset*: Simulated quasi-static potential due to tip-sample interaction. Only half of the tip is shown. **(C)** Moiré lattice in a tDBG. $\lambda$ denotes the moiré period. Red circles mark the ABBC stacking, while green and yellow indicate either ABAB or ABCA. **(D)** uMIM images of the moiré lattice in a tDBG with the magic-angle twist of ~1.3°. The stacking boundaries are superimposed onto the images, with the dots indicate the stacking following the color-code in (A). **(E)** uMIM signal profiles along the white dashed arrows in (D), averaged over 20 pixels width. The locations of different stacking are marked by colored dots. **(F)** A uMIM-Im image on a tDBG with isolated moiré defects. **(G)** The signal profile along the white arrow in (F).

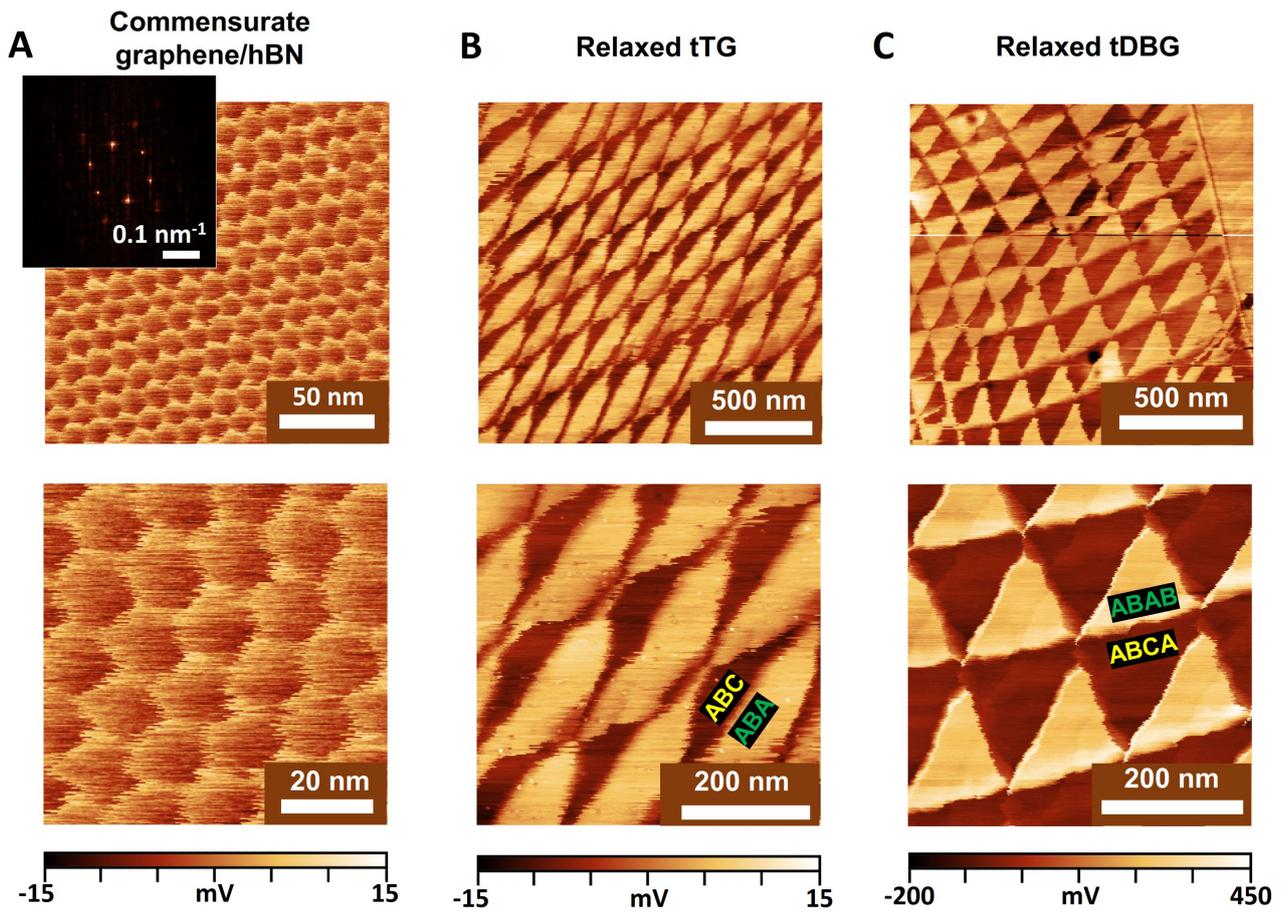

**Fig. 2. Versatility of uMIM in imaging various graphene-based moiré lattices.** **(A)** Commensurate, epitaxial monolayer graphene/hBN. **(B)** Near-0° tTG with relaxed ABA and ABC domains. **(C)** Near-0° tDBG with relaxed ABAB and ABCA domains. The upper row shows the large area scans of uMIM-Im signal. The FFT of (A) is shown as the inset. The lower row shows detailed uMIM-Im scans from each corresponding sample above.

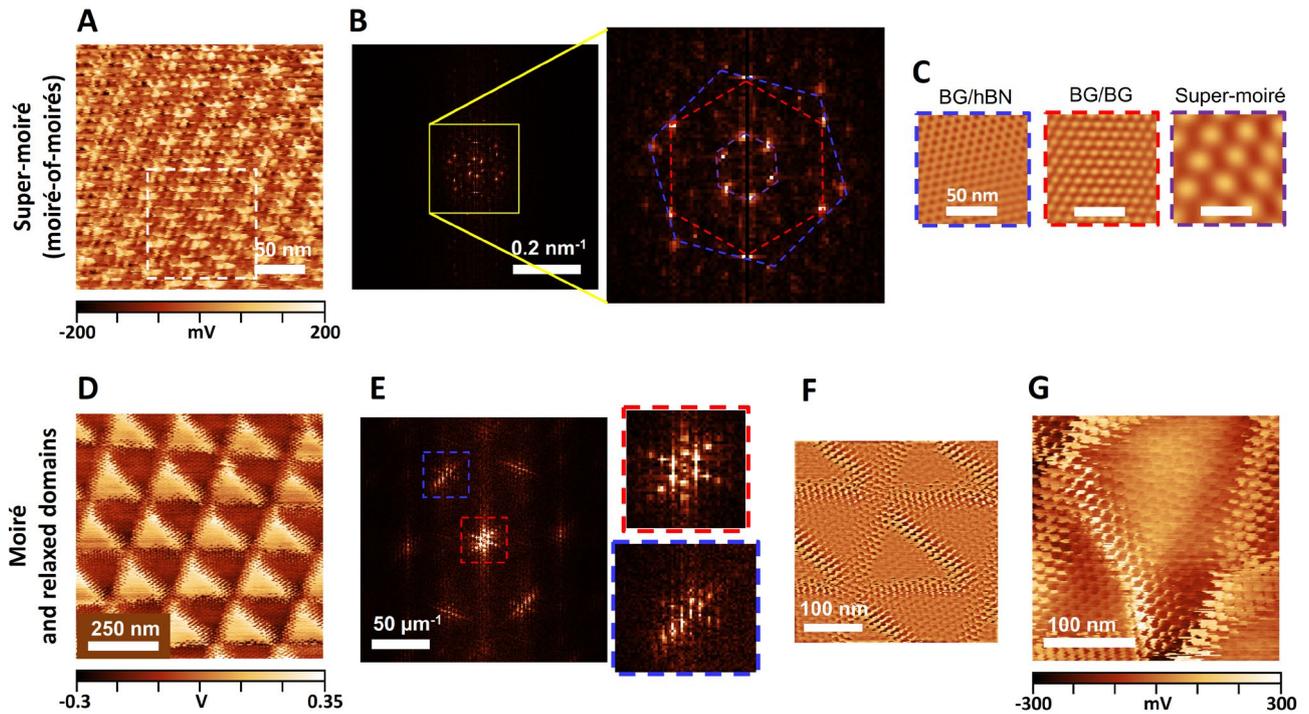

**Fig. 3. Superstructures from tDBG and hBN moirés. (A-C)** Super-moiré lattice: a moiré-of-moirés. (A) uMIM-Im image. (B) The FFT image of (A). The dashed hexagons marked the first-order period of lower BG/hBN moiré (blue), BG/BG moiré (red), and the emerging super-moiré (purple). (C) Fourier-filtered image of the area inside the white-dashed square in (A) based on the first-order moiré spots. **(D-G)** The composite of triangular ABAB-ABCA domains in near-0° tDBG with BG/hBN moiré. (D) uMIM-Im image. The BG/hBN moiré appear enhanced near the domain borders. (E) The FFT image of (D). The insets show the feature corresponding to the BG/hBN moiré (blue border), and triangular network (red border). (F) Fourier-filtered image of the features corresponding to the BG/hBN moiré. (G) Detailed image of a triangular domain.

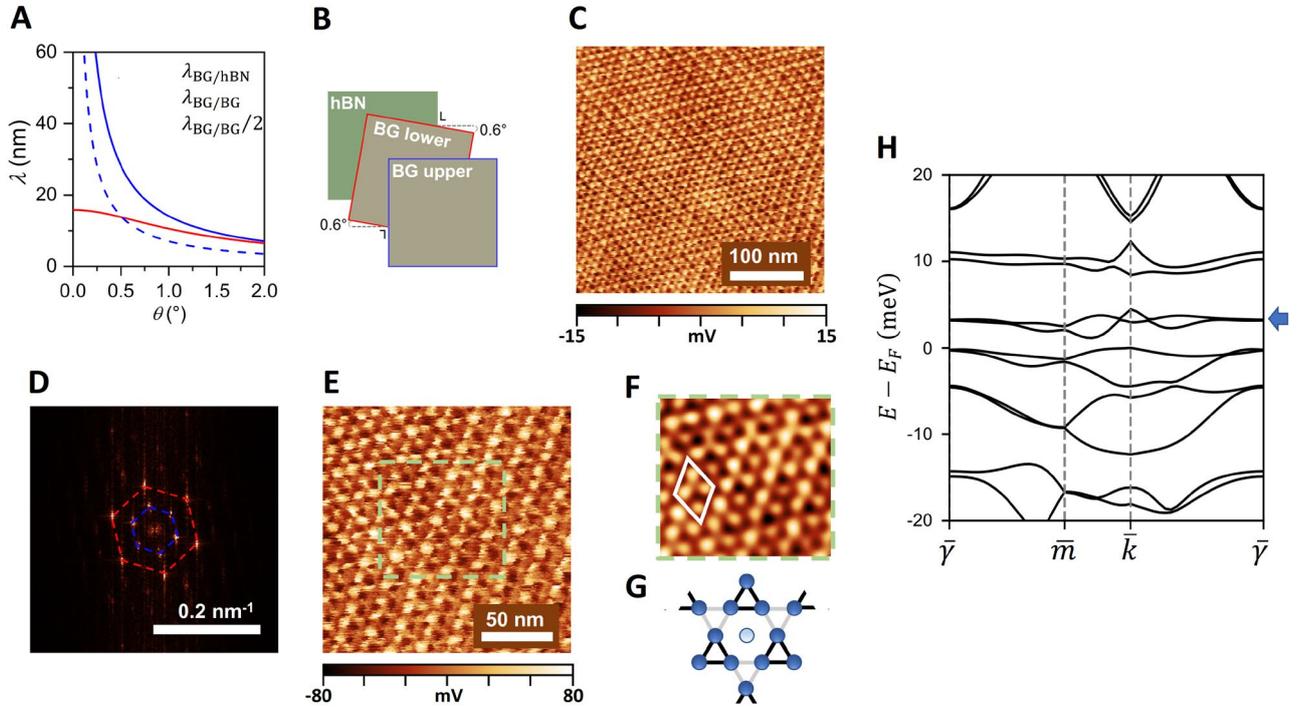

**Fig. 4. Kagome-like moiré superstructure in tDBG/hBN.** **(A)** Calculated moiré period of BG/BG and BG/hBN stacks as a function of twist angle. The condition $\lambda_{BG/BG}/\lambda_{BG/hBN} = 2$ is achieved at $\theta \approx 0.6°$. **(B)** The sample scheme to realize Kagome-like moiré. The BG/hBN and BG/BG flakes are twisted by 0.6°, but the hBN and the upper BG are aligned. **(C)** uMIM-Im image. **(D)** FFT of the image in (c). The dashed hexagons mark the first-order spots of BG/hBN moiré (red), and BG/BG moiré (blue). **(E)** Detailed uMIM-Im scan of Kagome-like moiré. **(F)** Low-pass filtered image from the area inside the green square in (E). The unit cell of the Kagome-like moiré is marked with white diamond. **(G)** An illustration of a trimerized Kagome lattice resembling the observed moiré. **(H)** Calculated band structure of the Kagome-like moiré lattice. The high symmetry points refer to that of the Brillouin zone of the BG/BG/hBN superstructure. The blue arrow marks the flat bands near the Fermi level.